\documentclass[%
 reprint,
 amsmath,amssymb,
 aps,
]{revtex4-2}

\usepackage{graphicx}
\usepackage{dcolumn}
\usepackage{bm}

\begin{document}

\preprint{APS/123-QED}

\title{Hybrid dielectrophoretic-optical trap for microparticles in aqueous suspension}

\author{Carlos D. Gonz\'alez-G\'omez,$^{a,b}$ Jose Garc\'ia-Guirado,$^c$ Romain Quidant,$^{c}$ F\'elix Carrique,$^{d,e}$ Emilio Ruiz-Reina,$^{b,e}$ and Raúl A. Rica-Alarc\'on$^{a,f,*}$}
\affiliation{
 $^{a}$Universidad de Granada, Department of Applied Physics, Nanoparticles Trapping Laboratory, 18071, Granada, Spain\\
$^{b}$Universidad de Málaga, Department of Applied Physics II, 29071, Málaga, Spain\\
$^{c}$ETH Zürich, Nanophotonic Systems Laboratory, 8092, Zürich, Switzerland\\
$^{d}$Universidad de Málaga, Department of Applied Physics I, 29071, Málaga, Spain\\
$^{e}$Institute Carlos I for Theoretical and Computational Physics (iC1), 29071, Málaga, Spain\\
$^{f}$Universidad de Granada, Research Unit “Modeling Nature” (MNat), 18071, Granada, Spain\\
$^*$E-mail:rul@ugr.es
}

\date{\today}

\begin{abstract}
We demonstrate that a set of microfabricated electrodes can be coupled to a commercial optical tweezers device, implementing a hybrid electro-optical trap with multiple functionalities for the manipulation of micro/nanoparticles in suspension. Our design allows us to simultaneously trap tens of particles in a single potential well generated in the low electric field region of the electrode arrangement, taking advantage of negative dielectrophoresis. Together with the optical tweezers, we show that the hybrid scheme allows enhanced manipulation capabilities, including controlled loading and accumulation in the dielectrophoretic trap from the optical tweezers, selectivity, and tracking of the individual trajectories of trapped particles.
\end{abstract}

\maketitle


\section{Introduction}

The capabilities to store, accumulate, separate, or filter micro/nanoparticles dispersed in a liquid are highly demanded for basic science and industry applications. Novel approaches that provide enhanced performance are continuously sought and developed, taking advantage of different physical principles and the distinguishing properties of the involved particles.\cite{lee2023review} Many of these functionalities are implemented in lab-on-a-chip devices for biomedical and chemical applications such as cancer diagnosis,\cite{Lambert2023Lab} detection of molecules,\cite{Xie2021Development} or detection of cells,\cite{Khan2020Ultrasensitive} to name but a few. 

During the last few decades, mature techniques such as optical tweezers have demonstrated their excellence in many applications,\cite{Gieseler2021Optical,Volpe2023Roadmap} including single-molecule and single-cell mechanics,\cite{bustamante2021optical} microrheology,\cite{robertson2018optical} sorting \cite{chapin2006automated}, as well as spectroscopic\cite{penders2018single,dai2021optical}  and non-equilibrium dynamics studies.\cite{Martinez2017Colloidal} The versatility of optical tweezers has favoured the development of commercial devices with different capabilities. Nonetheless, optical tweezers have some drawbacks. Heating occurs even in the case of dielectric particles,\cite{peterman2003laser,catala2017influence} leading to increased Brownian motion and instability,\cite{lu2021temperature,fernandez2023hot,ortiz2023light} and even damage when working with absorbing particles.\cite{Volpe2023Roadmap} Moreover, there are limitations regarding the optical properties of the used particles since their refractive index must be larger than that of the surrounding medium.\cite{spesyvtseva2016trapping} Another characteristic of optical trapping that can be a drawback for some applications lies in one of its strengths, i.e., they are intended to handle individual particles, not groups. One can typically trap a few particles in a single trap, but studying the dynamics of the individual particles in the trapping region would not be possible.\cite{Min2013Focused} In addition, optical tweezers are limited by diffraction phenomena, which prevents particles with sizes below a certain limit, typically of the order of 10 nm, from being trapped with standard techniques.\cite{juan2011plasmon}

Trapping can also be achieved using electric fields. A dielectrophoretic trap is a device that uses dielectrophoresis (DEP) to confine the particles. This phenomenon appears when particles move in the presence of an electric field gradient .\cite{zhang2010dielectrophoresis,pethig2010dielectrophoresis,d2017isolation} Depending on the dielectric properties of the particles and the electric field itself, DEP can be either positive (pDEP) or negative (nDEP), i.e. a particle can be attracted or repelled by the electrodes. DEP was comprehensively studied in the 50s by Herbert Pohl,\cite{Pohl1951Motion,Pohl1958Effects} and applications have been developed during the last few decades thanks to its versatility regarding manipulation, sorting, accumulation, exclusion, and trapping of particles with various sizes and properties. \cite{morgan2003ac,Rosenthal2005Dielectrophoretic,Lorenz2020High,Julius2023Dynamic,Turcan2020Dielectrophoretic,Russo2022Role}  

When particles are charged, Coulomb force can also lead to stable trapping in a device known as Paul or quadrupole traps.\cite{Paul1990Electromagnetic} In this case, stable trapping of charged particles can be obtained at a saddle point in an inhomogeneous AC electric field thanks to the appearance of a ponderomotive force. Interestingly, Paul traps can work either in liquid,\cite{Guan2011Paul} air,\cite{krieger2012exploring} or vacuum,\cite{wuerker1959electrodynamic}, and also need to be considered in the analysis of DEP, leading to rich dynamics.\cite{Guan2011Non,Park2012Stability}  

\begin{figure*}[htb]
	\centering
	\includegraphics[width=\linewidth]{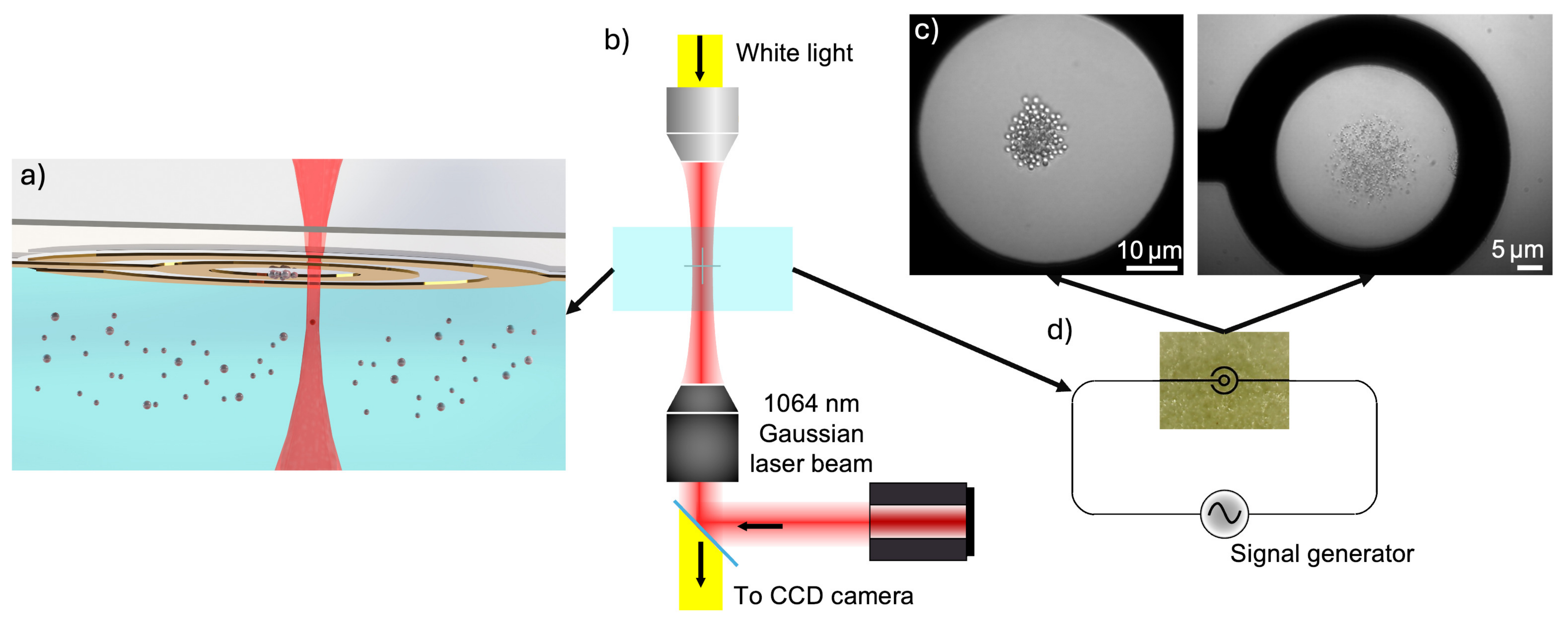}
	\caption{\textbf{Experimental setup of the hybrid DEP-optical trap.} a) Schematic representation of the experiment. A fluidic chamber about 100$\mu$m thick made by sandwiching two glass slides separated with high viscosity KORASILON\textsuperscript{\tiny\textregistered} paste contains an aqueous suspension of microparticles. Two concentrical ring microelectrodes (depicted in golden colour) lie on the top of the chamber, creating a dielectrophoretic trap that can confine several microparticles to a small volume. The DEP trap also generates an exclusion volume where no particles are allowed. The infrared beam (represented by the red beam) of the optical tweezers device can trap particles at its focus and can be used to bring new particles to the DEP trapping region, thanks to the capability of the device to steer the beam through all the field of the microscope. The setup can therefore be configured as DEP trap, optical trap, or hybrid DEP-optical trap, by switching on/off the electric field or the laser beam. b) The chip is introduced in a commercial optical tweezers platform (JPK-Bruker Nanotracker II), which consists of an inverted microscope with a tightly focused infrared laser beam ($\lambda$ = 1064 nm, 63x objective, NA = 1.2). c) Microscopy images of the central part of the microelectrodes illustrating the trapping of groups of 1~$\mu$m (left) and 500 nm (right) diameter polystyrene particles trapped in the centre of the inner electrode. We estimate a particle concentration of about 0.6 particles/$\mu\text{m}^2$ for the left image and about 1.2 particles/$\mu\text{m}^2$ for the right image, assuming in both cases they are arranged in a single plane. d) The microelectrodes are fed with a sinusoidal voltage signal with amplitudes of several tens of Volts at frequencies in the MHz range.}
	\label{fig:ExpSetUp}
\end{figure*}

Enhanced manipulation capabilities can be achieved by combining two trapping mechanisms.\cite{millen2015cavity,Conangla2018Motion,Conangla2020Extending,Bykov2022Hybrid} In these cases, Paul traps were used together with optical tweezers in a vacuum to reduce the laser power required to stably trap nanoparticles, and hence mitigate photodamage. Inspired by these developments and previous designs where circular electrodes were used to create both DEP\cite{Carlson2022DC,kwak2021size} and quadrupole traps,\cite{Alda2016Trapping} we here propose a hybrid trapping scheme that couples a set of microelectrodes to a commercial optical tweezers device (see Fig. \ref{fig:ExpSetUp}). We achieve DEP trapping of groups of microparticles in the same potential well. The video tracking capability of the microscope available in the optical tweezers device allows us to measure the dynamics of multiple particles together in the same potential well, which we illustrate by changing the number of trapped particles and modulating the trapping potential. Furthermore, we demonstrate selectivity in terms of the dielectric properties of the particles, as expected in DEP traps. Finally, we show that the optical tweezers allow one to selectively load the DEP trap from the optical tweezers by switching off the optical trap next to the DEP trap, and analyze the dynamics of a particle that simultaneously feels the two trapping mechanisms. 

The article is organized as follows. First, we provide a theoretical background about the dynamics of trapped particles and trapping mechanisms. Further, we present simulations of the electric field created by our microelectrodes arrangement that show the existence of a position that can stably trap particles with nDEP. We then present experimental evidence of the stability of the DEP trap, characterizing its performance for individual particles and then demonstrating some key capabilities, including selective trapping, control of the local density of trapped particles, and hybrid trapping, in particular demonstrating transfer of particles from the optical tweezers to the DEP trap. Finally, we finish with a summary of the results and a discussion of some potentialities of the presented device for further studies.

\section{Theoretical background}

\subsection{Dynamics of trapped particles}

Micro and nanoparticles dispersed in a fluid with viscosity $\eta$ at temperature $T$ are subjected to Brownian motion and diffusion, characterized by the diffusion coefficient $D$. The Einstein formula for the diffusion coefficient is $D=k_BT/\gamma$, where $k_B$ is the Boltzmann constant and $\gamma$ is the friction coefficient of the particle, which in the case of a sphere of radius $R$ far away from any walls can be written as $\gamma=6\pi\eta R$. Particle trapping techniques confine particles to a small region by applying a restoring force that pushes the particles to an equilibrium position. The restoring force can be of different origins, as we discuss below. The trapping mechanism prevents particles from diffusing away from an equilibrium position, given by the trap centre, but does not completely cancel Brownian fluctuations. Regardless of the trapping mechanism, the dynamics of Brownian particles dispersed in a liquid can be described by the overdamped Langevin equation: 

\begin{equation}
\label{eq:langevin}
    \gamma\frac{dx}{dt}+\kappa x=\sqrt{2k_BT\gamma\xi(t)}
\end{equation}

where $x$ is the instantaneous position of the particle, $\kappa$ is the trap stiffness, which can be due to diverse mechanisms, and $\xi(t)$ is a Gaussian white noise term that accounts for Brownian motion, with $<\xi(t)>=0$, $<\xi(t)\xi(t')>=\delta(t-t')$. The solution of this equation is well known, and its description of the dynamics of a trapped particle can be done considering the mean squared displacement (MSD) of the particle in the trap, which for the case of motion in a plane can be shown to be: \cite{Gieseler2021Optical}

\begin{equation}
\label{eq:BM1}
\text{MSD}(t)=4\frac{k_B T}{\kappa} \left(1-e^{-\frac{\kappa}{\gamma}t}\right)
\end{equation}

This expression shows the particle's motion has two regimes separated by a time scale $\tau=\gamma/\kappa)$. The particle diffuses freely at short times (MSD$(t<<\tau)\sim 4Dt$), while it features a plateau at long times (MSD$(t>>\tau) = 4k_BT/\kappa$), showing the trap's effect limiting the maximum excursions that it can execute away from the trap's centre. 

\subsection{Trapping mechanisms}

This section briefly describes the two trapping mechanisms used in this paper, namely optical tweezers and dielectrophoresis (see Fig. \ref{fig:ExpSetUp}). Interestingly, these two mechanisms are closely related and can be treated on similar grounds,\cite{riccardi2023electromagnetic} though we will discuss them separately. 

Firstly, a dielectrophoretic force is exerted by a non-uniform electric field on a polarisable particle. The time-averaged expression of this force in the dipole approximation is:\cite{Rosenthal2005Dielectrophoretic} 

\begin{equation}
\label{eq:DEP}
\textbf{F}_{\text{DEP}}=2\pi\varepsilon_{m}R^{3}\text{Re}\left[\underline{CM}(\omega)\cdot\nabla\underline{\textbf{E}}^{2}(\textbf{r},\omega)\right]
\end{equation}

where $\varepsilon_{m}$ is the electric permittivity of the medium where the particle is suspended, $\omega$ is the angular frequency of the applied electric field $\underline{\textbf{E}}$ and $\underline{CM}$ is the Clausius-Mossotti (CM) factor. For a bead, which can be considered a lossy dielectric uniform sphere, the expression of this factor is given by 

\begin{equation}
\underline{CM} (\omega)=\frac{\underline{\varepsilon}_{p}-\underline{\varepsilon}_{m}}{\underline{\varepsilon}_{p}+2\underline{\varepsilon}_{m}}
\end{equation}

where $\underline{\varepsilon}_{m}$ and $\underline{\varepsilon}_{p}$ are the complex electric permittivities of medium and particle, respectively. 

Regarding the CM factor, a particle can be attracted or repelled by the electric field gradient, i.e. the particle goes to the electrode, or away from it, respectively. This behaviour depends on the value of the CM factor. In many cases, each bead has one cut-off frequency, where the factor changes its sign from positive to negative. In the case of the polystyrene particles used in our experiments, the cut-off frequency is in the low MHz range. 

Secondly, an optical tweezers device is based on a single tightly focused light beam, which mainly exerts two optical forces namely the gradient and the scattering force. Let us briefly introduce the optical force experienced by an electric dipole in the presence of a time-varying electric field:\cite{Jones2015Optical}

\begin{equation}
\label{eq:OF}
    \textbf{F}_{\text{OT}}(\textbf{r})=\frac{1}{4}\alpha'_{\text{p}}\nabla\left|E\right|^{2}+\frac{\sigma_{\text{ext}}}{c}\textbf{S}
\end{equation}

where $E$ is the electric field amplitude, $\textbf{S}$ is the time-averaged Poynting vector of the incoming wave, $\sigma_{\text{ext}}$ is the extinction cross-section, $\alpha'_{\text{p}}=\text{Re}\left\{ \alpha_{\text{p}}\right\}$  is the real part of the polarisability and $c$ is the speed of light. The polarisability can be expressed as:\cite{Jones2015Optical}

\begin{equation}
    \alpha_{\text{p}}=\frac{\alpha_{\text{CM}}}{1-\frac{\varepsilon_{r}-1}{\varepsilon_{r}+2}\left[\left(k_{0}R\right)^{2}+\frac{2i}{3}\left(k_{0}R\right)^{3}\right]}
\end{equation}

where $\varepsilon_{r}$ is the relative electric permittivity and $k_{0}=2\pi/\lambda$. Also, $\alpha_{\text{CM}}=3V_s\varepsilon_{0}\frac{\varepsilon_{r}-1}{\varepsilon_{r}+2}$, where $\varepsilon_{0}$ is the electric permittivity of vacuum and $V_s=4/3\pi R^{3}$ is the volume of the sphere. 

The first term of Eq.~\ref{eq:OF} corresponds to the gradient force, which is responsible for confinement in optical tweezers. It is due to the potential energy of a dipole in the electric field and, thus, is a conservative force. On the other hand, the second term is related to the scattering force. This one is non-conservative and is produced by the transfer of momentum from the field to the particle. This is due to the scattering and absorption processes as revealed by the fact that it is proportional to the extinction cross-section $\sigma_{\text{ext}}$.\cite{Jones2015Optical}

In all situations discussed in this work, the forces given by Eqs.~\ref{eq:DEP} and ~\ref{eq:OF} can be assumed to follow Hooke's law in all three directions of space, and for every axis and force one finds a relation:
\begin{equation}
    F_{j,q}=-\kappa_{j,q}(q-q_{0,j})
\end{equation}
where $j\in \{\text{DEP,OT}\}$ indicates the considered trap, $q\in \{x,y,z\}$ are the coordinates of the particle, and $q_{0,j}$ are the corresponding locations of the equilibrium position of each trap.

\section{Results and discussion}

\subsection{COMSOL Multiphysics simulations}  
 
We performed electrostatics simulations with the COMSOL Multiphysics finite-elements platform to estimate the electric field created by our microelectrode configuration and help the interpretation of the experimental results. The simulation workflow is detailed in the Supplementary Information. 

Figure~\ref{fig:GradESim}a) shows the squared electric field distribution in the XY plane at a distance $z=10~\mu$m from the glass where the planar microelectrodes lie as obtained from simulations.  There, an electric field minimum can be observed in the centre of the electrodes at $(x,y)=(0,0)$, which leads to a stable dielectrophoretic trap if the CM factor is negative, in agreement with experiments. Panels b), c), and d) in Fig.~\ref{fig:GradESim} depict the gradient of the squared electric field for the Y-axis, X-axis, and Z-axis, respectively. Figure~\ref{fig:GradESim}d) shows that particles with negative CM factor are pushed towards the top wall above $z\simeq80\mu$m ($20~\mu$m from the plane where the electrodes lie, see Fig. S1). Experimentally, we observe that this force is enough to overcome gravity for the particles we tested. Moreover, the positive value of the gradient below the zero crossing leads to an exclusion region where no particles are allowed, as shown in panel d) and in Supplementary Video 1. This exclusion region is useful to realize clean experiments in the trap, so no undesired particles are allowed to accidentally enter the trapping region as they freely diffuse and/or are brought by convection that might be present in the chamber. 

\begin{figure*}[htb]
    \centering
    \includegraphics[width=\linewidth]{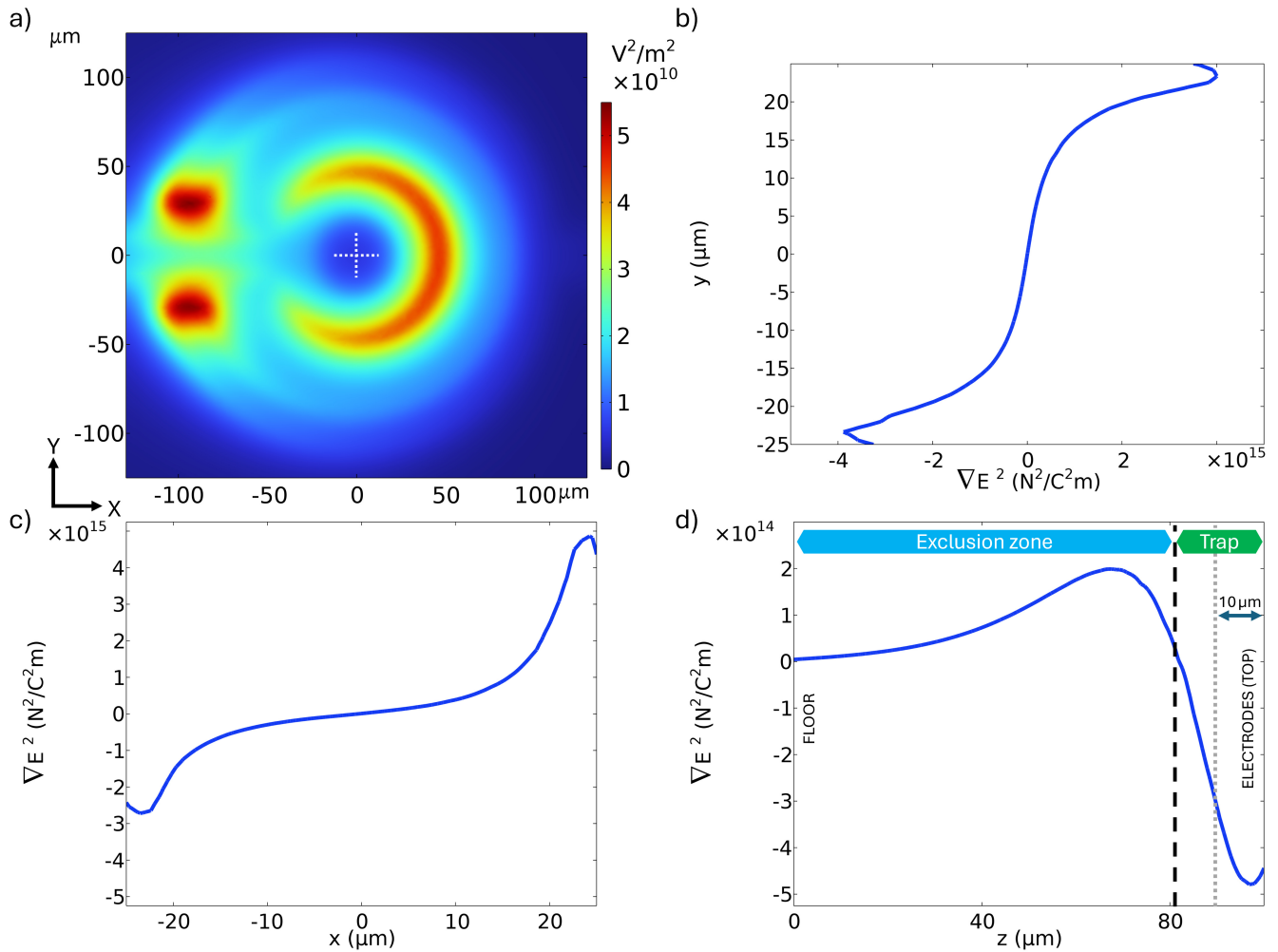}
    \caption{\textbf{COMSOL Multiphysics Simulation Results} a) Simulated squared electric field norm at 10$~\mu$m from the electrodes. Note the minimum in the centre of the electrodes, marked with a white dotted cross. Linear profiles of the  gradient of the squared electric field norm is also shown in  b) $x = 0$ and $z = 2~\mu$m, c) $y = 0$ and $z = 2~\mu$m and d) $x = 0$ and $y = 0$.}
    \label{fig:GradESim}
\end{figure*}

\subsection{Experimental results}
\subsubsection{Dielectrophoretic trap characterization}
The DEP trap presented here features a large influence volume, where particles that would otherwise freely diffuse are either pushed to the center of the trap or away from its influence region, generating an exclusion region. This can be expected from the field distribution seen in Fig. \ref{fig:GradESim}d), in agreement with experiments (see Supplementary Video 1). The exclusion region is even more apparent in the case of dense particles, since in our design the DEP trap is located on the top of the fluidic chamber, and sedimentation also contributes to the exclusion effect. 

We are interested in trapping at the centre of the ring (see Fig. \ref{fig:ExpSetUp}), away from electrodes, where video tracking of particles is possible. According to simulations, there is an electric field zero with a positive slope in the XY plane at this position, so we need to work in conditions where the CM factor is negative. This is typically achieved above a characteristic frequency $\omega_c\simeq (K_p+2K_m)/ (\varepsilon_p+2\varepsilon_m)$, where $K_p$ and $K_m$ are the conductivities of particles and medium, respectively.\cite{green1999dielectrophoresis} 

\begin{figure*}[ht!]
	\centering
	\includegraphics[width=\textwidth]{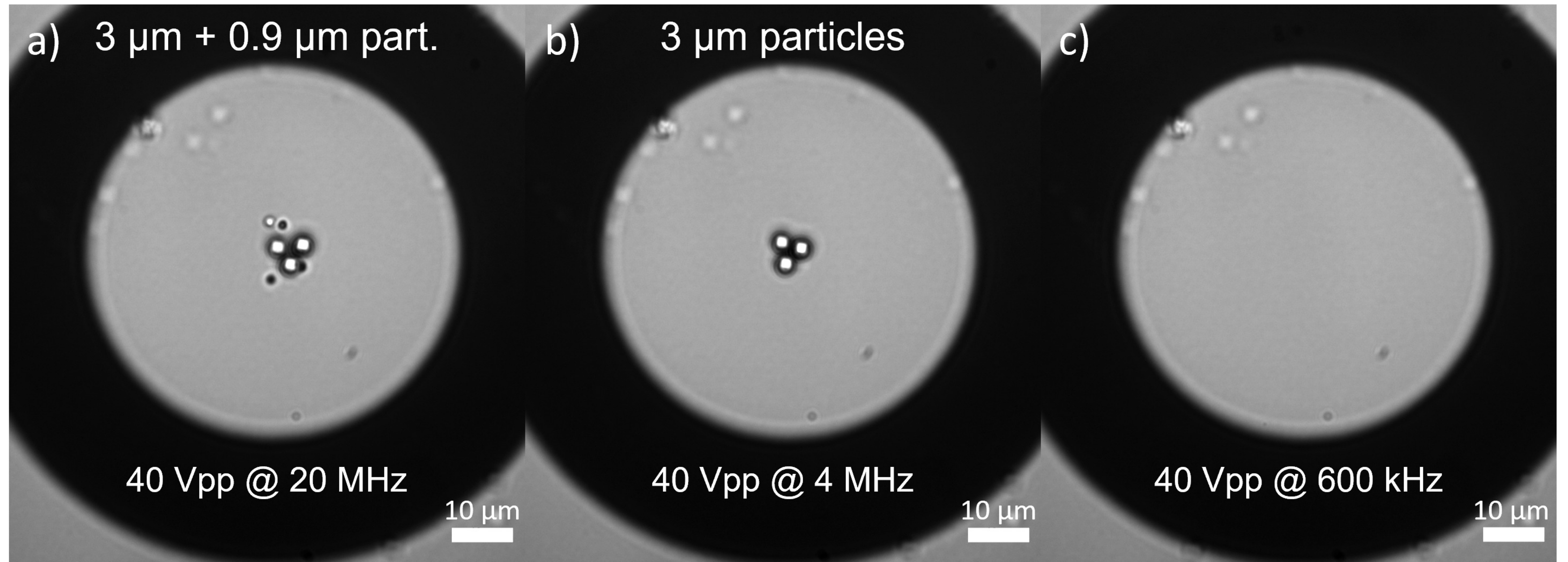}
	\caption{\textbf{Selective trapping}. a) The experiment starts with four 0.9$~\mu$m ProMag\textsuperscript{\tiny\textregistered} 1 particles and three $3.1~\mu$m ProMag\textsuperscript{\tiny\textregistered} 3 particles trapped with an electric field of 40~Vpp and 20~MHz. b) At a given time, we changed the frequency to 4~MHz, kicking out the small particles. c) Larger particles can also be expelled from the trap by reducing the frequency down to 600 kHz.}
	\label{fig:PartFilter}
\end{figure*}
Dielectrophoretic traps are well known for their selective stability based on the cut-off frequency in the CM factor. To demonstrate this capability in our device, we introduced in a microchamber a mixture of 0.9$~\mu$m and 3.1$~\mu$m ProMag\textsuperscript{\tiny\textregistered} particles. We used particles with a clear difference in radius to highlight clearly their different behavior. As can be seen in panel a) of Fig. \ref{fig:PartFilter} (see also Supplementary Video 2), we observe that at an electric voltage signal of 40 $\text{V}_{\text{pp}}$  and a frequency of 20 MHz both large and small particles are trapped. When we reduce the frequency to 4 MHz (panel b)), the smaller particles are expelled from the trap. Further, pushing out the larger particles is possible, reducing the frequency even more to 600 kHz (panel c)). This last ejection can also be achieved by switching OFF the electric field. However, in this case, the particles leave the trap mainly by diffusion and sedimentation, while in the previous case, the particles are expelled by the positive DEP force.  

We evaluated the characteristic frequency for different particles in the DEP trapping device. To do so, we selected particles from different sizes and materials, including polystyrene, ProMag\textsuperscript{\tiny\textregistered} (a type of composite particles made of a mixture of polymeric matrix with embedded magnetite), and SiO$_2$ particles (see Materials and Methods). Our results are shown in Table~\ref{tab:cutoff}, demonstrating that this parameter can be used as a way to selectively push particles out of the trap. In particular, the cutoff frequency decreases with particle size, in agreement with previous observations.\cite{green1999dielectrophoresis} A detailed analysis of how the cutoff frequency depends on particle properties is out of the scope of the present work. 

\begin{figure}[ht!]
	\centering
	\includegraphics[width=\linewidth]{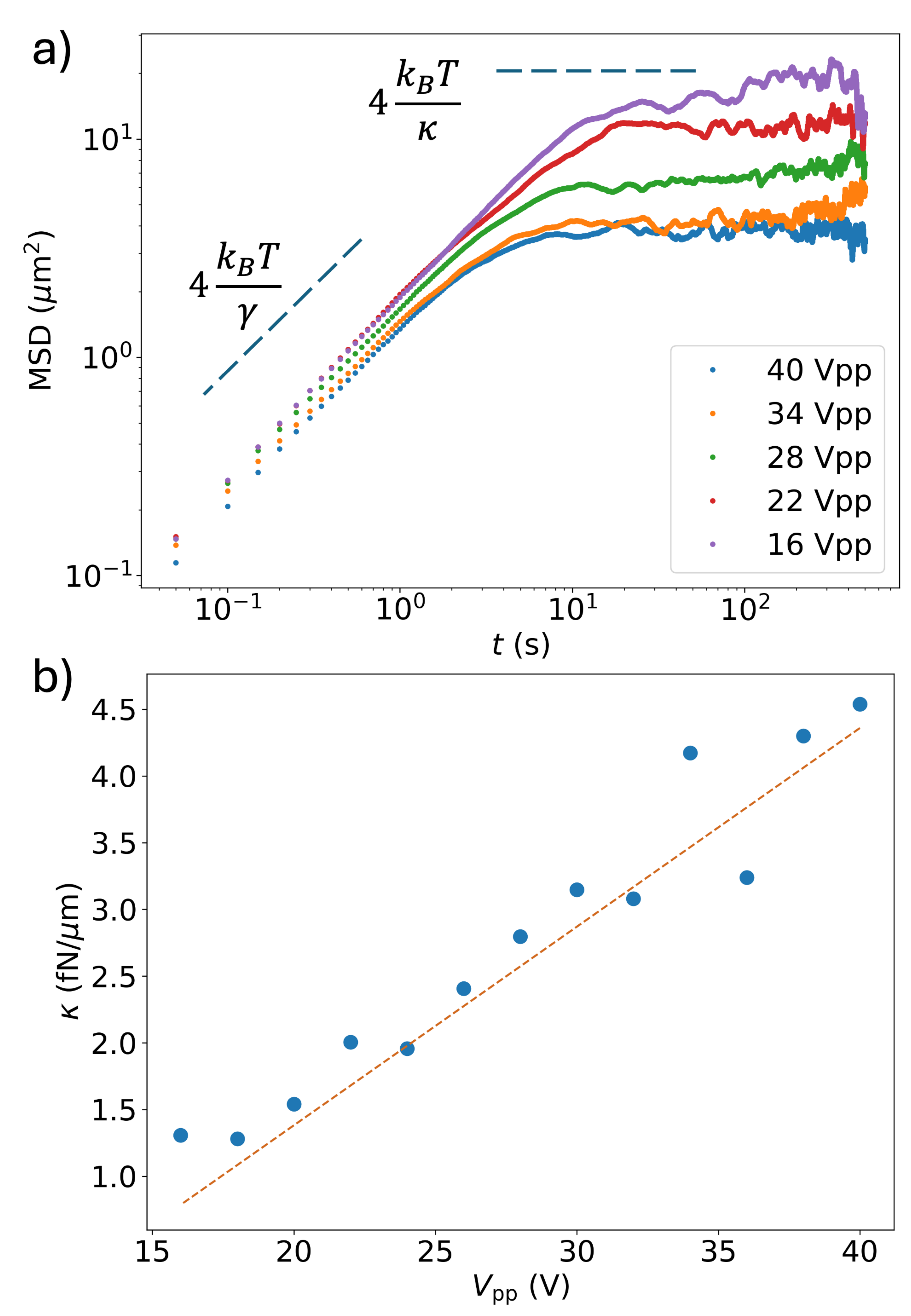}
	\caption{\textbf{Evaluation of the trapping performance of individual particles.} a) Mean squared displacement for different voltages for ProMag\textsuperscript{\tiny\textregistered} 1 particles with the electric field frequency fixed at 10 MHz. Notice that the particles behave as freely diffusing for short times with the slope $4k_BT/\gamma$, while featuring a plateau at $4k_BT/\kappa$ above the characteristic time defined by $\gamma/\kappa$.  b) Trap stiffness $\kappa$ for different voltage values. The dashed line corresponds to a linear fit with slope 0.15$\pm$0.01 $\text{fN}/(\text{V}\mu \text{m})$ and intercept -1.60$\pm$0.28 $\text{fN}/\mu \text{m}$. } 

	\label{fig:single_characterization}
\end{figure}

We characterized the trapping performance for a single absorbing microparticle (ProMag\textsuperscript{\tiny\textregistered} 1, mean diameter $0.906~\mu$m) in the DEP device, obtaining the results shown in Fig.~\ref{fig:single_characterization}. Due to their strong absorption, these particles are hard to trap in an optical tweezers device.\cite{siler2014anomalous,quinto2014microscopic,fernandez2023hot} To that aim, we recorded videos of the Brownian motion of a single particle in the trap and computed the MSD of the trajectories obtained with Trackpy, which is a particle tracking package for Python (see Materials and Methods).~\cite{Allan2018trackpy,Vanrossum1991Interactively} The motion corresponds to quasi-2D diffusion in a parabolic potential, as shown in Fig.~\ref{fig:single_characterization}a). The MSD is proportional to time for short times, while it features a plateau above a characteristic time, in agreement with Eq.~\ref{eq:BM1}. We observe that the confinement increases as the voltage is raised, and for a fixed $T\simeq 300$~K, this means an increase in the stiffness of the trap. Even if our device does not have precise temperature control, the fact that the slope of the MSD at short times does not change significantly ($\sim$12\% between 16 and 40 $V_{\text{pp}}$) with voltage indicates that the temperature was moderately stable throughout the experiment, and the observed variation in the plateau level cannot be explained by temperature changes. From fits of experimental data to Eq.~\ref{eq:BM1}, we obtain values for the stiffness of the trap $\kappa$ and the friction coefficient $\gamma$. As shown in Fig.~\ref{fig:single_characterization}b), the stiffness increases linearly with the applied voltage. In contrast, the friction coefficient did not show any clear dependence, obtaining $\gamma=1.02\pm 0.25 \times 10^{-8}$~Ns/m, where the uncertainty refers to the standard error of the mean. This value is larger than the expected from the Stokes friction coefficient $\gamma\simeq0.73\times 10^{-8}$~Ns/m at $T=300$~K. This disagreement can be solved by considering that the trapping occurs very close to the top of the chamber and that the friction coefficient should be corrected for the hydrodynamic interactions.\cite{Schaffer2007Surface} Using Faxen's law, the dependence of $\gamma$ with the distance to the wall can be evaluated, obtaining that trapping occurs at a distance of about 0.4~$\mu$m to the wall, in agreement with the fact that we observe that trapping occurs slightly displaced from the plane where the electrodes are seen in focus.

\begin{table}[]
\begin{center}
\begin{tabular}{l|ccc|clc|cc|}
\cline{2-9}
& \multicolumn{3}{c|}{Polystyrene (PS)}                           & \multicolumn{3}{c|}{ProMag\textsuperscript{\tiny\textregistered}}          & \multicolumn{2}{c|}{Silica (Si)}     \\ \hline
\multicolumn{1}{|l|}{Reference}                                                     & \multicolumn{1}{c|}{200} & \multicolumn{1}{c|}{500} & 1000 & \multicolumn{2}{c|}{1} & 3      & \multicolumn{1}{c|}{500} & 1000 \\ \hline
\multicolumn{1}{|l|}{$f_c$ (MHz)} & \multicolumn{1}{c|}{10}  & \multicolumn{1}{c|}{7}   & 3    & \multicolumn{2}{c|}{9}   & 2         & \multicolumn{1}{c|}{5}   & 2    \\ \hline
\end{tabular}
\caption{\textbf{Cutoff frequency for different particles.} Frequency below which trapping is not possible at the electric field minimum due to a change to positive CM factor, i.e., the frequency at which there is 
 a transition from nDEP to pDEP.} 

\label{tab:cutoff}
\end{center}
\end{table}

\subsubsection{Tracking the Brownian dynamics of groups of particles in the DEP trap}
 
\begin{figure}[ht!]
	\centering
	\includegraphics[width=\linewidth]{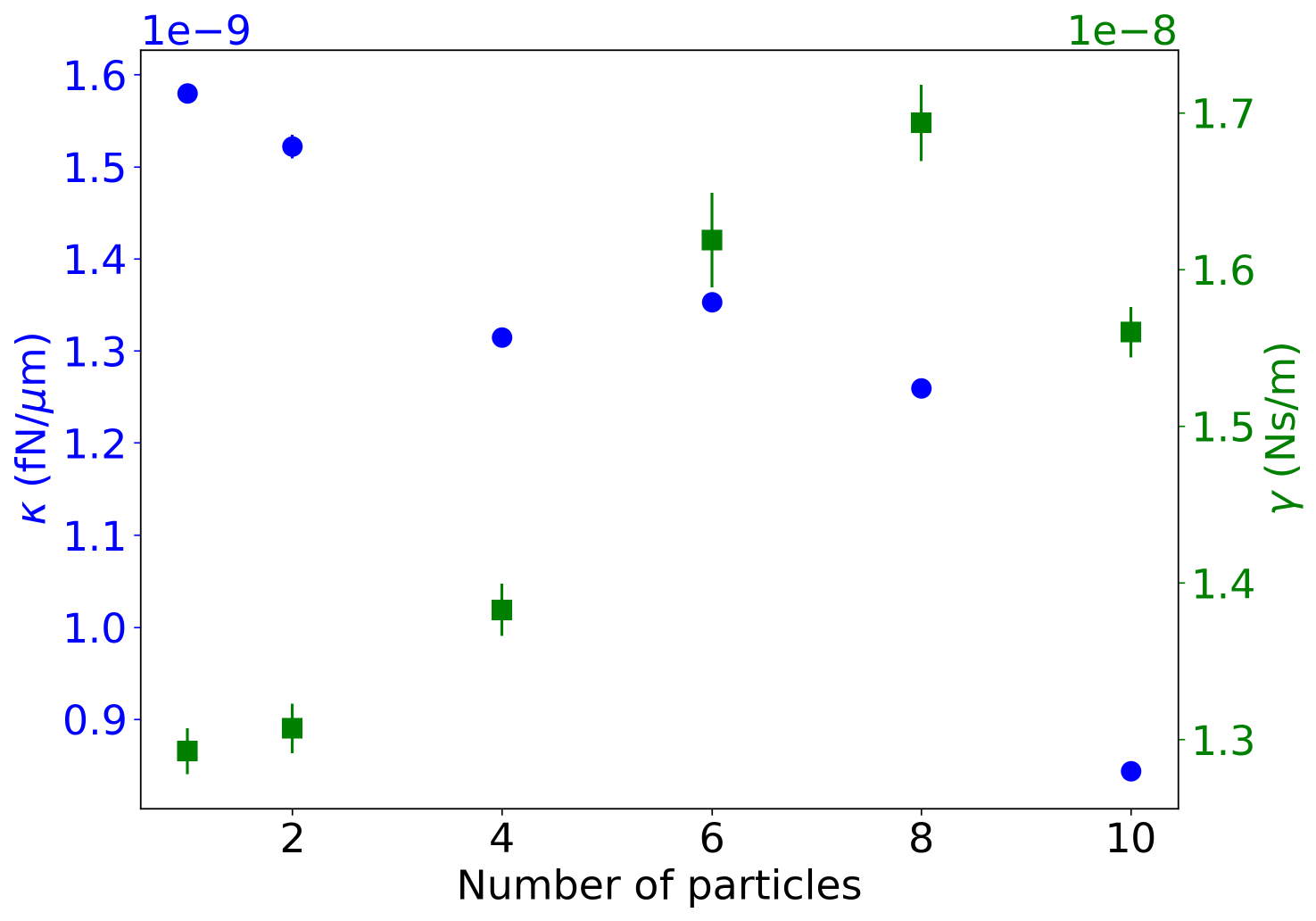}
	\caption{\textbf{Characterization of the effect of multiple  ProMag\textsuperscript{\tiny\textregistered} 1 particles in the DEP trap}. a) Trap stiffness (left axis) and viscous friction coefficient (right axis) for different groups of ProMag\textsuperscript{\tiny\textregistered} 1  particles. Error bars stand for the uncertainty of the fit.}   
	\label{fig:Multiparticle}
\end{figure}

A key capability of the DEP trap, as opposed to conventional optical tweezers, consists of trapping several particles while tracking their trajectories individually. The information obtained this way can be used, e.g., to analyse how interactions between particles affect their Brownian dynamics inside a parabolic potential. To illustrate this capability, we recorded videos with increasing numbers of ProMag\textsuperscript{\tiny\textregistered} 1 particles in the trap at 40 $\text{V}_{\text{pp}}$ and $f$=10 MHz, conditions that we had previously found to lead to stable trapping situation. After that, the trajectories were extracted for every particle in each video using a modified Trackpy code.\cite{Allan2018trackpy} Further, we calculated for every experiment the individual mean square displacement (IMSD) and the ensemble mean square displacement (EMSD), i.e., the average of all of the individuals in each group. We analysed experiments with up to 10 particles simultaneously stored in the trap since our tracking code did not provide reliable results for larger groups of particles. The results of these experiments are summarized in Fig.~\ref{fig:Multiparticle}. Increasing the number of particles in the trap leads to a decrease in stiffness and an increase in the friction coefficient. However, these results should be considered with caution. Even if an increase in the friction coefficient is expected because of particle-particle interactions, the observed decreasing trend in trap stiffness is not justified and is likewise due to interactions unaccounted in our description based on Eqs.~\ref{eq:langevin} and ~\ref{eq:BM1}, which consider a single particle in a parabolic potential. 

\subsubsection{DEP Trap Modulation}

\begin{figure}[ht!]
	\centering
	\includegraphics[width=\linewidth]{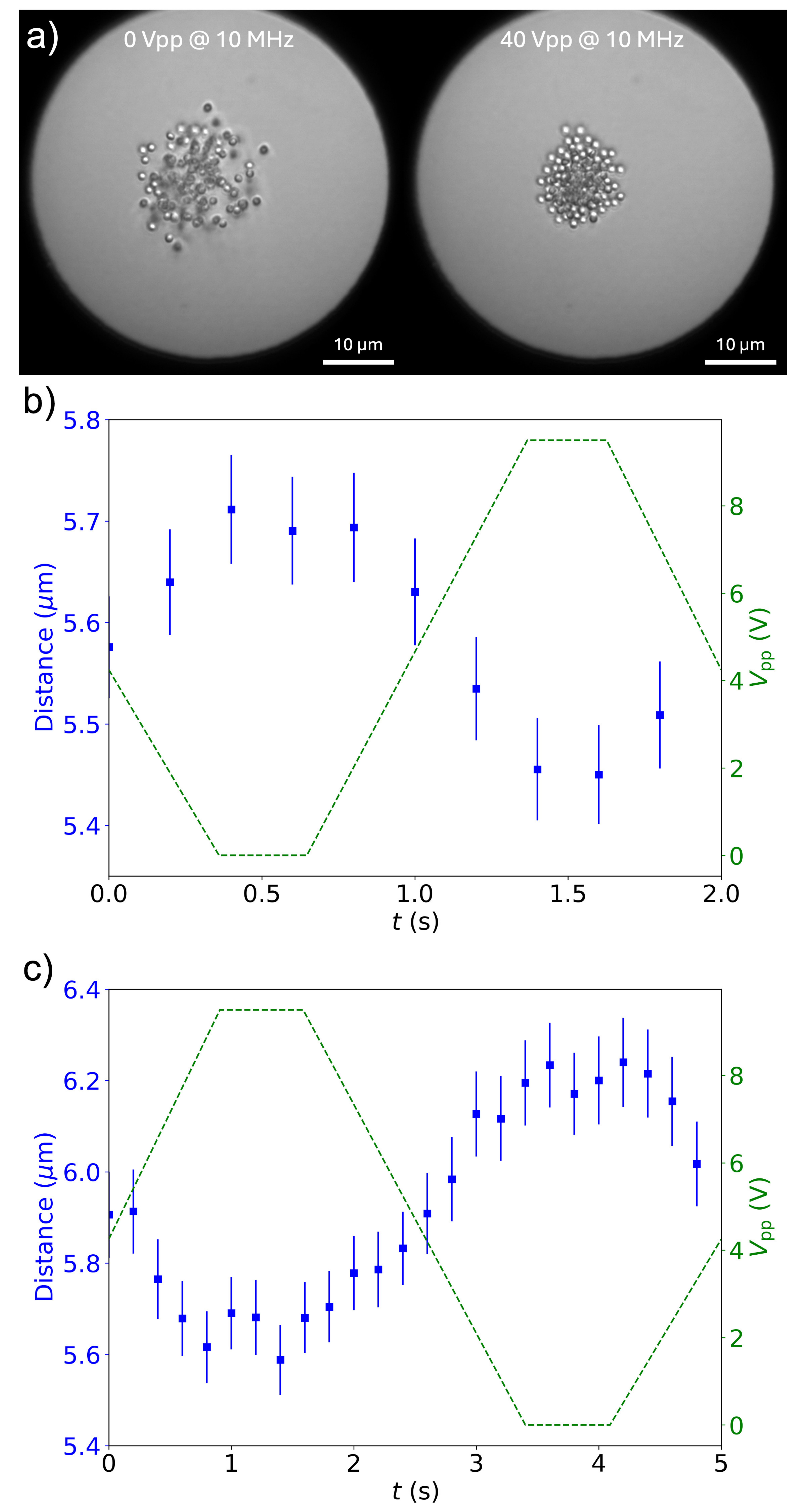}
	\caption{\textbf{Trap Modulation}. a) Cycles of expansion (left subpanel) and compression (right subpanel) of 1~$\mu$m polystyrene particles. Averaged distance between the 1~$\mu$m polystyrene particles for a 2- (b)) and 5-s (c)) period cycles, respectively. The green dashed line is the modulated amplitude of the feed voltage, while the blue dots are the averaged distance between every two particles, vs cycle time. Error bars stand for the standard deviation from the repeated cycles. In c), the left inset corresponds to compression (higher electric field), while the right one corresponds to expansion (lower electric field).}
	\label{fig:TrapMod}
\end{figure}

The trap stiffness can be controlled through the voltage applied to the electrodes. When several particles are trapped together, modulating the voltage also modulates the trap stiffness, effectively changing the local density of particles, as shown in Fig.~\ref{fig:TrapMod}. In panel a), we illustrate the effect of voltage modulation with some tens of 1~$\mu$m polystyrene particles (see also Supplementary Videos 3 and 4). When the voltage is low (below ~4 V, see Fig.~\ref{fig:TrapMod}a) left), the particles explore a large volume due to Brownian fluctuations. The stiffness increases as the voltage is raised (above ~4 V) and the particles are strongly pushed towards the trap's centre, as seen in the right image in Fig.~\ref{fig:TrapMod}a). 

Quantitative experiments of density modulation can be done with a small number of particles. We performed two experiments where the amplitude of the electric field was modulated according to the protocols shown in Fig.~\ref{fig:TrapMod}b) and c), corresponding to modulation cycles of period 2~s and 5~s, respectively. We chose two periods arbitrarily but different enough to understand how the frequency affects the behavior of the system. We analyzed this type of experiment by recording videos at 5 fps of the trajectories of 16 (600 cycles of modulation at 500 mHz, i.e., cycles of period 2~s) and 14 particles (240 cycles of modulation at 200 mHz, i.e., cycles of period 5~s). These trajectories were analyzed with Trackpy, and the distance between every two particles was measured and averaged out at every time frame. Finally, further averaging over cycles leads to the results shown in Fig.~\ref{fig:TrapMod}b) and c). The average distance between particles decreases when the amplitude of the 10 MHz signal is maximum, while it decreases in the parts of the cycle when the trap is relaxed and even switched off. Notice that the particles do not move only in the focal plane (XY) but are also displaced along the optical axis (Z), being defocused when the electric field is low and getting back into focus when the electric field is high. Unfortunately, our tracking software could not account for displacements along this axis, and the data provided corresponds to the projection on the XY plane. Although during some parts of the period, the voltage is almost zero, the particles did not escape because the diffusion time of the particles is greater than the time-lapse with 0 V, and the particles are recaptured periodically. 

\subsubsection{Hybrid Trap}

\begin{figure}[ht!]
	\centering
	\includegraphics[width=\linewidth]{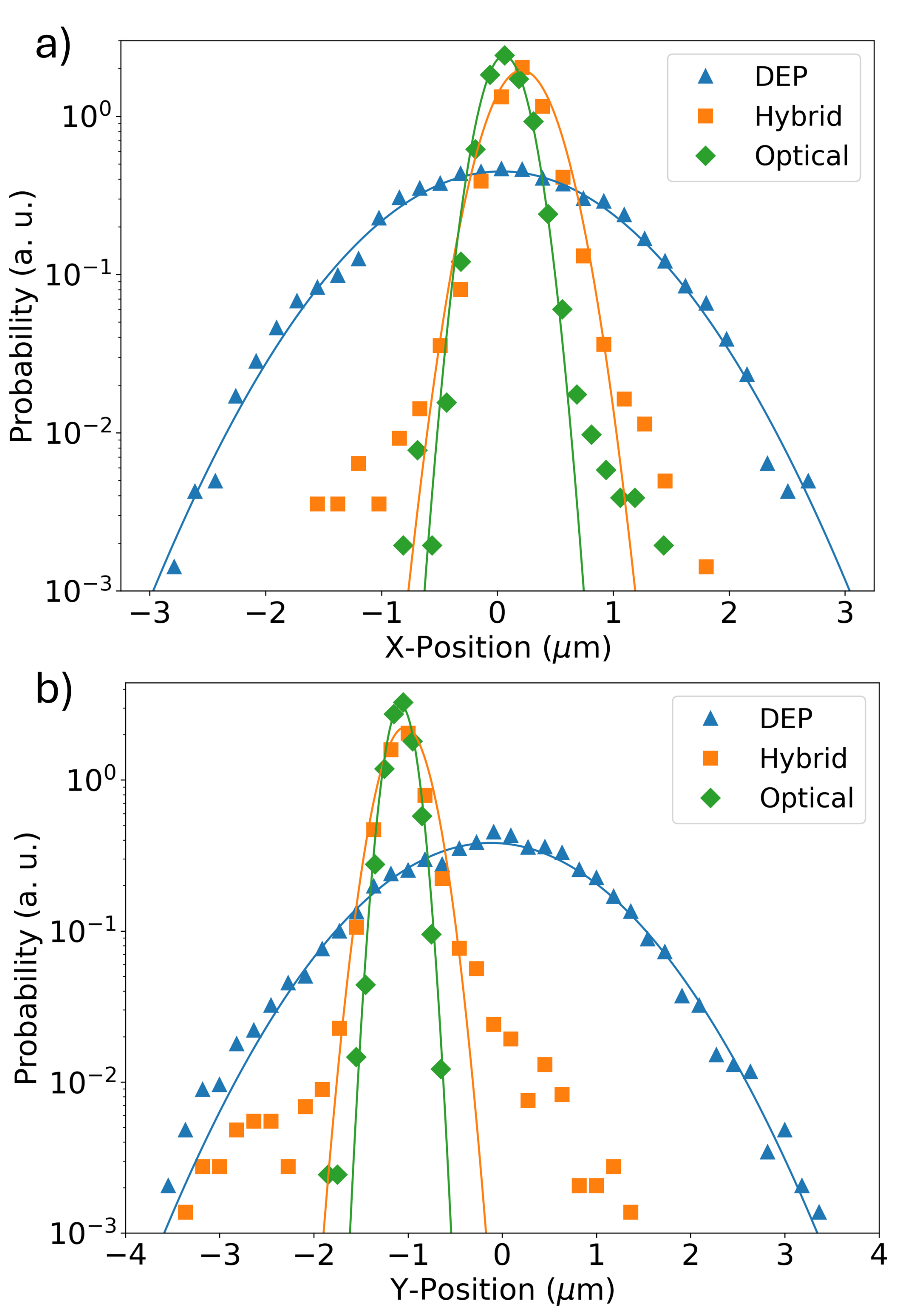}
	\caption{\textbf{Hybrid DEP-optical trap}. 1D Histograms of a) x-position and b) y-position of a $1~\mu$m polystyrene particle. Blue triangles correspond to DEP trap, green diamonds to the optical trap and orange squares to the combined trap. Solid lines represent fits to Gaussian functions. The power of the laser beam was the same (1~mW) both in the experiments performed with the optical and hybrid traps.} 

	\label{fig:HybridTraps}
\end{figure}

One of the most interesting features of our design is the possibility it offers to be combined with optical tweezers, leading to a hybrid trap with enhanced manipulation capabilities.\cite{Conangla2020Extending,Bykov2022Hybrid,bonvin2023hybrid} We here demonstrate that optical tweezers are compatible with the DEP trap, and illustrate how these two approaches can be used to, e.g., load the DEP trap with particles previously trapped in optical tweezers. In the first experiment, we trapped one 1$\mu$m-polystyrene particle with the optical trap (see Fig. \ref{fig:ExpSetUp} and Materials and Methods) using a power of 1~mW and then brought it to the vicinity of the electrodes when the electric signal was switched off. After recording one passive trajectory, we obtained the position histograms depicted with green diamonds in Fig.~\ref{fig:HybridTraps}. As can be seen, the histogram thus obtained is Gaussian, and from its variance and Equipartition Theorem, we obtain estimations of the stiffness along the X-Y axes in different configurations. The stiffness of the optical trap (OT) in the X-axis is $\kappa_{\text{OT},x}\simeq1.4\cdot10^{-7}$ N/m and $\kappa_{\text{OT},y}\simeq1.1\cdot10^{-7}$ N/m in the Y-axis. As the laser power was very small, the optical trap alone was not stable enough to hold the particle for a long time. Interestingly, the hybrid trap was stable with this low power in conditions when DEP trap was stable. In this case, the particle did not escape from the optical trap but remained trapped for some minutes. Analyzing the trajectory of this experiment returns the histogram shown in orange squares in Fig.~\ref{fig:HybridTraps}. In this case, one can observe non-Gaussian tails due to the presence of the electric field, but this did not destabilize the trap. The stiffness of the hybrid trap (HT) is slightly reduced, obtaining $\kappa_{\text{HT},x}\simeq7\cdot10^{-8}$ N/m and $\kappa_{\text{HT},y}\simeq5\cdot10^{-8}$ N/m, but similar to that obtained only with OT. Finally, the particle was captured by the DEP trap when we switched off the optical trap. In this case, the histogram of the position is significantly broader (blue triangles in the figure), returning smaller values of the stiffness ($\kappa_{\text{DEP},x}\simeq6\cdot10^{-9}$ N/m and $\kappa_{\text{DEP},y}\simeq4\cdot10^{-9}$ N/m). Under typical working conditions, the stiffness obtained with optical tweezers is higher than those achievable with DEP traps, though the potential depth of the latter is larger. 

\begin{figure*}[ht!]
	\centering
	\includegraphics[width=\linewidth]{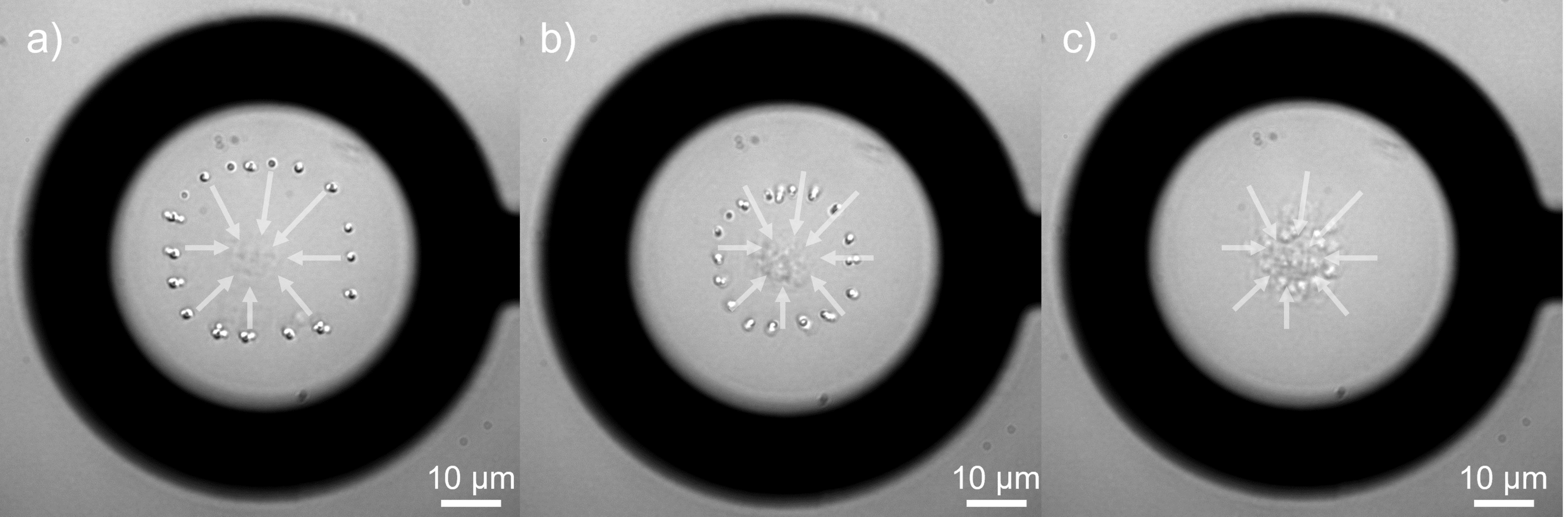}
	\caption{\textbf{Controlled loading of the DEP trap from optical tweezers} a) $1~\mu$m $\text{SiO}_2$ particles optically trapped within the DEP trap. b) and c) After we switched off the optical traps, the particles quickly moved to the centre of the DEP trap. White arrows indicate the net displacement of particles throughout the video. }     
	\label{fig:HybridTrapsRing}
\end{figure*}

Beyond demonstrating the hybrid trap, we here present an example of the manipulation capabilities of our hybrid trap. In this case, as shown in Fig.~\ref{fig:HybridTrapsRing} (see also Supplementary Video 5), we optically trapped a few 1$~\mu$m $\text{SiO}_{2}$ particles arranged in a circumference thanks to an acousto-optic deflector that allows the optical tweezers to work in time-sharing mode,\cite{pesce2020optical,Gieseler2021Optical} thus effectively generating multiple traps (panel a)). The acousto-optic deflector is controlled with the software included in the commercial optical tweezers device. In the same image, the presence of particles stored in the DEP trap can be seen at a different plane (see the electrodes are out of focus). Moments later, we switched off the optical traps, so the particles moved towards the DEP trap (panel b)). Finally, in the last frame (panel c)), all the particles are inside the trapping region of the DEP trap. The hybrid trapping with the exclusion region discussed earlier allows one to realize experiments in clean conditions since no particles can accidentally enter the DEP trapping region. 

\section{Conclusions and outlook}
We have designed and characterized a novel DEP trap that can store microparticles in aqueous suspension. The microelectrodes arrangement has broad optical access, so it can be combined with an optical tweezers platform, allowing for a hybrid trapping scheme. We have shown that a single particle in the DEP trap behaves like an overdamped Brownian particle in a parabolic potential, similar to a particle in optical tweezers. Interestingly, the DEP trap is not limited to dielectric particles but sets the stage for experiments with different materials (e.g., carbon, metallic, or other types of light-absorbing particles). 

Significantly, we have also demonstrated that the DEP trap allows one to store several Brownian particles in the same potential well and simultaneously track their dynamics either in equilibrium or under the modulation of the trapping potential. This opens the door to the implementation of paradigmatic experiments in statistical physics that have so far only been performed with individual particles in optical tweezers, including fundamental questions in stochastic thermodynamics\cite{toyabe2010experimental,berut2012experimental,ciliberto2017experiments} implementation of stochastic heat engines\cite{blickle2012realization,Martinez2016Brownian,Martinez2017Colloidal} or counterintuitive relaxation phenomena,\cite{martinez2016engineered,Kumar2020Exponentially,pires2023optimal,ibanez2024heating}, to the case of multiple interacting particles. 

Finally, the demonstrated capabilities of selective trapping and hybrid operation permit the realization of applications where sorting, filtering, or selective loading can be needed during the implementation of biochemical assays, as the interaction of different species could be forced or modulated for sensing applications.

\section{Acknowledgements}
Financial support from FEDER/Junta de Andalucía-Consejería de Economía y Conocimiento/Project P18-FR-3583, Grant PID2021-127427NB-I00 funded by MICIU/AEI/10.13039/501100011033, and by ERDF "A way of making Europe" is acknowledged. 

\bibliographystyle{elsarticle-num}


\end{document}